\documentclass[11pt]{article}
\usepackage{latexsym}
\usepackage{graphics}
\usepackage{epsfig}

\newcommand{\bt}{\begin{tabbing}}
\newcommand{\et}{\end{tabbing}}

\newcommand{\ord} {{\cal R}}
\newcommand{\p} {{\prime}}
\newcommand{\pp} {{\prime\prime}}

\newtheorem{definition}{Definition}
\newtheorem{theorem}{Theorem}
\newtheorem{lemma}{Lemma}

\newtheorem{corollary}{Corollary}

\newtheorem{claim}{Claim}
\newcommand{\qed}{\rule{2mm}{3mm}}

\begin{document}

\title{Finding a sun in building-free graphs}

\author{
Elaine M. Eschen
\thanks{
elaine.eschen@mail.wvu.edu, Lane Department of Computer Science and
Electrical Engineering, West Virginia University, Morgantown, WV
26506. Acknowledges support from NSF WV EPSCoR. }
\and
Ch\'{\i}nh T.
Ho\`ang
\thanks{
choang@wlu.ca, Department of Physics and Computer Science, Wilfrid
Laurier University, Waterloo, ON N2L 3C5, Canada. Acknowledges
support from NSERC of Canada. }
\and
Jeremy P. Spinrad
\thanks{spin@vuse.vanderbilt.edu, Department of Electrical Engineering and
Computer Science, Vanderbilt University, Nashville, TN 37235.}
\and
R. Sritharan
\thanks{
srithara@notes.udayton.edu, Computer Science Department, The
University of Dayton, Dayton, OH 45469. Acknowledges support from
The National Security Agency, USA.} }

\maketitle

\begin{abstract}
Deciding whether an arbitrary graph contains a sun was recently
shown to be NP-complete \cite{H}. We show that whether a
building-free graph contains a sun can be decided in O(min$\{m{n^3},
m^{1.5}n^2\}$) time and, if a sun exists, it can be found in the
same time bound. The class of building-free graphs contains many
interesting classes of perfect graphs such as Meyniel graphs which,
in turn, contains classes such as hhd-free graphs, i-triangulated
graphs, and parity graphs. Moreover, there are imperfect graphs that
are building-free. The class of building-free graphs generalizes
several classes of graphs for which an efficient test for the
presence of a sun is known. We also present a vertex elimination
scheme for the class of (building, gem)-free graphs. The class of
(building, gem)-free graphs is a generalization of the class of
distance hereditary graphs and a restriction of the class of
(building, sun)-free graphs.
\end{abstract}

\section{Introduction}
For a fixed graph $\mathcal{F}$, we say a graph $G$ is
$\mathcal{F}$-free if it does not contain $\mathcal{F}$ as an
induced subgraph. For a set $\mathcal{S}$ of graphs, we say a graph
$G$ is $\mathcal{S}$-free if it does not contain any member of
$\mathcal{S}$ as an induced subgraph.  When we say ``a graph
contains $\mathcal{F}$" (``a graph does not contain $\mathcal{F}$"),
we mean that the graph contains (does not contain) $\mathcal{F}$ as
an induced subgraph.

A {\it house} is the complement of an induced path on five vertices.
A {\it hole} is an induced cycle on five or more vertices. A {\it
domino} is the graph with vertex set $ \{a, b, c, d, e, f \} $ and
edges $ab$, $bc$, $cd$, $de$, $ef$, $fa$, and $be$. A graph is {\it
chordal} if it does not contain any induced cycle on four or more
vertices. A {\it sun} is a chordal graph with the Hamiltonian cycle
$(x_1, y_1, x_2, y_2,\ldots ,$ $ x_n, y_n)$, $n \geq 3$, in which
each $x_i$ is of degree exactly two and the $y_i$ vertices form a
clique.

The problem of deciding whether a graph contains a sun was recently
shown to be NP-complete \cite{H}. However, algorithms exist that can
find a sun in polynomial time provided the input graph belongs to a
restricted class. A graph is {\it strongly chordal} if it is chordal
and does not contain a sun \cite{F}. As chordal graphs can be
recognized in linear time \cite{RTL} and as strongly chordal graphs
can be recognized in O(min$\{n^2, m\log{n}\}$) time \cite{PT,S}, it
follows that a sun in a chordal graph can be found in O(min$\{n^2,
m\log{n}\}$) time. A graph is {\it hhd-free} if it does not contain
a house, hole, or domino as an induced subgraph. The class of
hhd-free graphs properly generalizes the class of chordal graphs.
The class of hhd-free graphs was originally studied in the context
of perfectly orderable graphs \cite{C}. Whether a given graph is
hhd-free can be determined in O($n^3$) time \cite{HS,g1}. A graph is
{\it hhds-free} if it is hhd-free and does not contain a sun. It was
shown recently that the class of hhds-free graphs can be recognized
in O($n^3$) time \cite{EHS} (the first polynomial-time algorithm for
the problem appears in \cite{g2}). Thus, deciding whether a given
hhd-free graph contains a sun can be done in O($n^3$) time. In light
of the hardness of testing for a sun in general \cite{H},
determining for which classes of graphs one can test for the
presence of suns efficiently is interesting.

In this paper, we show that whether a building-free graph contains a
sun can be tested in O($m{n^3})$ time. A {\it building} is the graph
obtained from a hole by adding exactly one edge $e$ so that the edge
$e$ forms a triangle with two incident edges of the hole. Thus, a
house is a building on five vertices. Our algorithm can also find a
sun in a given building-free graph, should it exist, in O($m{n^3}$)
time. It is easily seen that every hhds-free graph is building-free.
In fact, every Meyniel graph is building-free; a graph is a {\it
Meyniel graph} if every odd cycle with at least five vertices in the
graph has at least two chords. The class of Meyniel graphs is
perfect and contains such classes of graphs as hhd-free,
i-triangulated, and parity. The class of building-free graphs is
rather large and it is not contained in the class of perfect graphs.
For example, a hole on odd number of vertices is building-free, but
is imperfect.

A {\it gem} is the graph with vertex set $\{a, b, c, d, e\}$ where
$\{a, b, c, d\}$ induces a $P_4$ and $e$ is adjacent to each of $a$,
$b$, $c$, and $d$.  We show that every (building, gem)-free graph
admits a certain elimination scheme of vertices. The class of
(building, gem)-free graphs is a generalization of the class of
distance hereditary graphs; {\it distance hereditary graphs} are
exactly those graphs that are (house, hole, domino, gem)-free. It is
seen that every sun contains a gem. Thus, the class of (building,
gem)-free graphs is contained in the class of (building, sun)-free
graphs.

Our algorithm borrows ideas used in \cite{EHS} for the recognition
of hhds-free graphs. However, in the interest of completeness, we
provide full details in our paper, while clearly noting how the
ideas from \cite{EHS} are used. The algorithm in \cite{EHS}, in a
sense, reduces the problem of finding a sun in the input graph to
finding a sun on six vertices in a derived chordal graph. Deriving
such an auxiliary chordal graph and looking for a sun on six
vertices in that graph involve properties of strongly chordal graphs
and the process of ``strongly chordal completion'', which is
discussed later. In view of this, we next present some known
properties of strongly chordal graphs. First, we need some
definitions.

We use $N(x)$ to denote the set of vertices adjacent to vertex $x$
in a graph $G$; i.e., $N(x)$ is the set of {\it neighbours} of $x$.
The set $N(x)$ is referred to as the {\it (open) neighbourhood of
x}, while the set $N[x] = N(x) \cup \{x\}$ is the {\it closed
neighbourhood of x}. The set of {\it nonneighbours of x} is denoted
by $M(x) = V(G) - N[x]$.  For vertices $x$ and $y$ of graph $G$, we
say $x$ {\it sees} $y$ when $x$ and $y$ are adjacent in $G$, and $x$
{\it misses} $y$ when $x$ and $y$ are not adjacent in $G$.

Vertex $x$ in a graph is {\it simplicial} if $N(x)$ induces a
complete graph. It is well known \cite{D} that graph $G$ is chordal
if and only if every induced subgraph $H$ of $G$ contains a
simplicial vertex of $H$. Farber \cite{F} proved an analogous
characterization for the class of strongly chordal graphs. Vertex
$x$ in a graph is {\it simple} if the vertices in $N(x)$ can be
ordered as ${x_1},{x_2}, \ldots, {x_k}$ such that $N[{x_1}]
\subseteq N[{x_2}] \subseteq \ldots \subseteq N[{x_k}]$. Thus, every
simple vertex is simplicial. For a graph $G$, let ${\cal R} =
{v_1},{v_2}, \ldots, {v_n}$ be an ordering of vertices of $G$. Let
$G(i)~=~G[\{{v_i}, {v_{i+1}}, \ldots, {v_n}\}]$; i.e., $G(i)$ is the
subgraph induced in $G$ by the set of vertices $\{{v_i}, {v_{i+1}},
\ldots, {v_n}\}$. ${\cal R}$ is a {\it simple elimination ordering}
for $G$ if $v_i$ is simple in $G(i)$, $1 \leq i \leq n$. ${\cal R}$
is a {\it strong elimination ordering} for $G$ if (i) ${\cal R}$ is
a simple elimination ordering, and (ii) for every $i < j < k$ such
that $v_j, v_k \in N(v_i)$, $N[{v_j}] \subseteq N[{v_k}]$ in $G(i)$.
In other words, in a strong elimination ordering, for any vertex
$v_i$, the neighbours of $v_i$ in $G(i)$ appear in ${\cal R}$
according to the order of inclusion of their closed neighbourhoods
in $G(i)$. The following is due to Farber \cite{F}:
\begin{theorem}[\cite{F}]
The following are equivalent for any graph $G$:
\begin{itemize}
\item $G$ is strongly chordal. \item $G$ is chordal and does not
contain a sun. \item Vertices of $G$ admit a simple elimination
ordering. \item Vertices of $G$ admit a strong elimination
ordering.
\end{itemize}
\end{theorem}

\subsection{The algorithm}
Note that whether an arbitrary graph contains a building can be
decided in O($m{n^3}$) time. For every vertex $v$ of the graph, for
every pair $x$, $y$ of adjacent vertices in $N(v)$, we do the
following: delete all the other neighbours of $v$, delete the vertex
$v$, delete every vertex in $N(x) \cap N(y)$, delete the edge $xy$,
and check if a path connecting $x$ and $y$ exists in the remaining
graph. Thus, though we present our algorithm in such a way that it
takes a building-free graph as input and tests for the presence of a
sun, it can easily be modified so that it takes an arbitrary graph
as input and tests for the presence of a sun or a building in the
graph. We need some definitions first.

\begin{definition}\label{def:sun}
By a {\it k-sun} $((d_1,d_2, \ldots, d_k),( c_1, c_2, \ldots, c_k))$
with $k \geq 3$ we denote the graph obtained by taking a clique on
vertices $c_1, c_2, \ldots, c_k$,  a stable set on vertices
$d_1,d_2, \ldots, d_k$, and for each $i$, $1 \leq i \leq k$, adding
edges $d_i c_i, d_i c_{i+1}$ with the subscripts taken modulo $k$.
The vertices $d_i$ are called the {\it tips} of the sun.  For an
arbitrary $k$, we refer to a $k$-sun simply as a sun.
\end{definition}

\begin{definition} \label{def:sunflower} {\normalfont{\cite{EHS}}}
By a {\it sunflower} $((P_1,P_2, \ldots, P_k),( c_1, c_2, \ldots,
c_k))$ with $k \geq 3$ we denote the graph obtained by taking an
arbitrary graph with vertices $c_1, c_2, \ldots, c_k$,  and for each
$i$, $1 \leq i \leq k$, adding an induced path $P_i$ of length at
least two connecting $c_i$ to $c_{i+1}$ with the subscripts taken
modulo $k$. The paths $P_i$ are called the petals of the sunflower.
The edges of the petals are called boundary edges of the sunflower.
Vertices $c_i$ are called center vertices, the remaining vertices of
the sunflower are called petal vertices.  If a path $P_i$ has length
two, the petal vertex of $P_i$ is called a tip of the sunflower.
\end{definition}

First we explain the basic logic behind the algorithm. Then, we
present the algorithm and its proof of correctness. Suppose $G'$ is
a building-free graph. If a sun exists in $G'$, then some vertex $x$
must be a tip of a sun in $G'$, and since a sun is a sunflower, $x$
is also a tip of a sunflower in $G'$. Further, the only two
neighbours of $x$ in this sunflower are adjacent. We will show later
that if a vertex $x$ is a tip of a sunflower $S$ in $G'$ such that
the only two neighbours of $x$ in $S$ are adjacent, then a sun must
exist in $G'$. Therefore, our algorithm checks every vertex $x$ of
$G'$ to determine whether $x$ is such a tip of a sunflower in $G'$.
In order to do this, for every edge $yz$ in the neighbourhood of
$x$, we consider the graph $G$ obtained from $G'$ by deleting all
the neighbours of $x$ except $y$ and $z$, and determine whether $x$
is the tip of a sunflower in $G$. Note that $G$ is building-free,
vertex $x$ has only two neighbours in $G$, and $x$ is simplicial in
$G$.

\bt xxxx\=xxxx\=xxxx\=xxxx\=xxxx\kill
{\bf Algorithm} {\it find-sun}\\
{\bf Input}:  building-free graph $G'$\\
{\bf Output}: {\it true} if $G'$ contains a sun; {\it false} otherwise\\
\\
\{\\
\> {\bf for} each vertex $x$ of $G'$ {\bf do}\\
\> \> {\bf for} each edge $yz$ of $G'$ in $N(x)$ {\bf do}\\
\> \> \{ \\
\> \> \> Obtain graph $G$ from $G'$ by deleting all the neighbours
of $x$\\
\> \> \> except $y$ and $z$. \\
\> \> \> {\bf if} ({\it tip-of-sunflower}($G$, $x$)) {\bf then}\\
\> \> \> \>        return ({\it true});\\
\> \> \{ \\
\> return ({\it false}); \\
\} \et

Next, we present algorithm {\it tip-of-sunflower}, which was used in
the context of hhd-free graphs in \cite{EHS}. We will show that it
can be adopted for use in the context of building-free graphs also.

 \bt
xx\=xxxx\=xxxx\=xxxx\=xxxx\=xxxx\kill
{\bf Algorithm} {\it tip-of-sunflower}\\
{\bf Input}:  building-free graph $G$ and a simplicial vertex $x$ of $G$\\
{\bf Output}: {\it true} if $x$ is the tip of a sunflower in $G$; {\it false} otherwise\\
\\
\{\\
\>1.\> For each vertex $y \in M(x)$ compute $n(y,x) = |N(y) \cap N(x)|$.\\
\\
\>2.\> Sort $M(x)$ into $\ord$ $= {y_1},{y_2}, \ldots, {y_k}$ \\
\> \> such that
           $i < j$ implies $n(y_{i},x) \leq n(y_{j},x)$.\\
\\
\>3.\> /* Perform strongly chordal completion of $G[M(x)]$. */\\
\>\> {\it SCC} ($G[M(x)]$);\\
\>\> Obtain graph $H$ from $G$ by adding to $G$ the edges that were\\
\>\> added to $G[M(x)]$ during the strongly chordal completion of $G[M(x)]$.\\
\>\> /* Now $\ord$ is a strong elimination ordering of $H[M(x)]$. */\\
\\
\>4.\> {\bf if} ({\it in-3-sun}($H$, $x$)) {\bf then}\\
\> \> \>         return ({\it true})\\
\>\> {\bf else}\\
\> \> \>         return ({\it false});\\
\}
\et

Now we prove some facts about the algorithm {\it tip-of-sunflower}.
For the sake of proving the facts about the algorithm, we will
assume that strongly chordal completion of $G[M(x)]$ in step~3 of
Algorithm {\it tip-of-sunflower} is performed as in the following
$k$ iterations (as given in \cite{EHS}):

 \bt xxxx\=xxxx\=xxxx\=xxxx\=xxxx\=xxxx\=xxxx\=xxxx\kill
{\bf Algorithm} {\it SCC} \\
{\bf Input}:  Graph $G=(V,E)$ and an arbitrary ordering $\ord$ of $V$\\
{\bf Output}: Graph $G_k =(V, E_k)$ where $E \subseteq E_k$ and \\
$\ord$ is a strong elimination ordering for $G_k$. \\
\{\\
1. $G_0$ = $G$ \\
2. {\bf for} $i$ = 1 {\bf to} $k$ {\bf do}\\

\> // \> We make $y_i$  simple in $G_i[\{{y_i},{y_{i+1}}, \ldots, {y_k} \} ]$ such that for $i < p < q$ \\
\> // \> the following holds with respect to  $G_i[\{{y_i},{y_{i+1}}, \ldots, {y_k} \} ] $:\\
\> // \> if $y_p \in N({y_i})$ and $y_q \in N({y_i})$, then $N[{y_p}] \subseteq N[{y_q}]$.\\

\> Obtain $G_i$ via updating $G_{i-1}$ by adding edges as
follows: \\
\> \> Let $A= N_{G_{i-1}}(y_i) \cap \{{y_{i+1}}, \ldots,
{y_k}\}$. \\
\>  \> 2.1 \>  Form $G_i^1$ by adding to $G_{i-1}$ edges between the vertices of $A$\\
\> \> \>  so that $A$ is a clique.\\
\> \> 2.2 \> Let the vertices of $A$ be $y_{i_1}, y_{i_2,} \ldots,y_{i_t}$, where $i_1 < \ldots < i_t$. \\
\> \> \> Let $G_i^2 = G_i^1$.
\\
\> \> \> {\bf for} $j = 1$  {\bf to} $t-1$ {\bf do}\\
\> \> \> \> For each pair of vertices $y_r$, $y_{i_s}$, \\
\> \> \> \> where $y_r \in \{y_{i+1}, \ldots, y_k\}$
and $y_{i_s} \in A, j+1 \leq s \leq t$,\\
\> \> \> \> such that in $G_{i-1}$, $y_{i_j}$ sees $y_r$ and $y_{i_s}$ misses $y_r$,\\
\> \> \> \> add edge $y_ry_{i_s}$ to  $G_i^2$.\\
\> \> 2.3 \> $G_i = G_i^2$ \\
\}
 \et

Strongly chordal completion, as described in {\it SCC}, requires
O($n^3$) time. It was shown in \cite{EHS} that it can be implemented
to run in O($n^2$) time.

\begin{definition}\label{def:dom} {\normalfont{\cite{EHS}}}
Given a graph $G=(V,E)$, and a set $X \subseteq V$, we say that
vertex $u$ $X$-dominates vertex $v$ if $N(v) \cap X \subseteq N[u]
\cap X$. We say $u$ dominates $v$ to mean $u$ $V$-dominates $v$.
\end{definition}

Let $M(x) = \{y_1, y_2, \ldots, y_k\}$ and $\ord$ be the ordering
of the vertices of $M(x)$ produced by step~2 of Algorithm {\it
tip-of-sunflower}. For convenience, we write $y_i < y_j$ when $i <
j$. Note that whether or not a vertex $N(x)$-dominates another
vertex is unaffected by the additions of edges as a result of the
strongly chordal completion of $G[M(x)]$, as these edges are
always between vertices of $M(x)$.

Lemma~\ref{lem:pathdom} was proved in \cite{EHS} for (house,
hole)-free graphs. We show it holds for the more general class of
building-free graphs as well.

\begin{lemma} \label{lem:pathdom}
Let $G$ be a building-free graph with simplicial vertex $x$. Let
$y^\prime, y^{\prime\prime} \in M(x)$ be such that $y^\prime$ comes
before $y^{\prime\prime}$ in $\ord$. If $G$ has an induced path $P$
connecting $y^\prime$ and $y^\pp$ all of whose vertices are in
$M(x)$ and such that each internal vertex of $P$ is $N(x)$-dominated
by $y^\p$ or $y^\pp$, then $y^\pp$ $N(x)$-dominates $y^\p$ (and
hence, $y^\pp$ $N(x)$-dominates each vertex of $P$).
\end{lemma}
\noindent {\bf Proof of Lemma~\ref{lem:pathdom}.} Suppose $G$ has an
induced path $P$ that satisfies the hypothesis, but $y^\pp$ does not
$N(x)$-dominate $y^\p$. Then $y^\p$ sees $a \in N(x)$ that $y^\pp$
misses; also, since $n(y^\p, x) \leq n(y^\pp, x)$, $y^\pp$ sees $b
\in N(x)$ that $y^\p$ misses. No internal vertex of $P$ sees both
$a$ and $b$, since each such vertex is $N(x)$-dominated by $y^\p$ or
$y^\pp$. In a traversal of $P$ starting at $y^\p$, let $y_b$ be the
first vertex of $P$ that sees $b$. Vertex $y_b$ exists since $y^\pp$
sees $b$. In a traversal of $P$ from $y_b$ towards $y^\p$, let $y_a$
be the first vertex that sees $a$.  Vertex $y_a$ exists since $y^\p$
sees $a$. Now either there are no vertices between $y_a$ and $y_b$
on $P$ and $\{x,a,b,y_a,y_b\}$ induces a house in $G$, or $\{x,
a,b,y_a,y_b\}$ along with the vertices on $P$ between $y_a$ and
$y_b$ (all of which must miss both $a$ and $b$) induces a building
in $G$. In either case, we have a contradiction. Thus, $y^\pp$
$N(x)$-dominates $y^\p$ and, by transitivity, $y^\pp$
$N(x)$-dominates every vertex on $P$. \hfill $\qed$

Lemma~\ref{lem:hchordal} through Lemma~\ref{lem:flower} were proved
in \cite{EHS} in the context of (house, hole)-free graphs. Given
Lemma~\ref{lem:pathdom} (for building-free graphs), the proofs of
these lemmata given in \cite{EHS} are valid without modification for
building-free graphs. Rather than reproduce the proofs here, we
refer the reader to \cite{EHS}.

\begin{lemma} \label{lem:hchordal}
Let $G$ be a building-free graph with simplicial vertex $x$. Let $H$
be the graph as constructed in Algorithm {\it tip-of-sunflower}. Let
${x_1}, \ldots, {x_r}$ be an arbitrary ordering of vertices in
$N(x)$. Then, $H$ is chordal with \\ ${y_1}, \ldots,
{y_k},{x_1},\ldots, {x_r},x$ as its perfect elimination scheme.
\end{lemma}

\begin{lemma} \label{lem:sun}
Let $G$ be a building-free graph with simplicial vertex $x$. Let $H$
be the graph as constructed in Algorithm {\it tip-of-sunflower}. If
vertex $x$ is in a sun in $G$, then $x$ is in a 3-sun in $H$.
\end{lemma}

\begin{lemma} \label{lem:flower}
Let $G$ be a building-free graph with simplicial vertex $x$. Let $H$
be the graph as constructed in Algorithm {\it tip-of-sunflower}. If
vertex $x$ is in a 3-sun in $H$, then $x$ is in a sunflower in $G$.
\end{lemma}

\begin{lemma} \label{lem:sunflower}
Let $G$ be a building-free graph. Suppose $G$ contains a sunflower
$S$ such that some petal $P$ of $S$ has exactly two edges and the
only petal vertex of $P$ is simplicial in $S$. Then, $G$ contains a
sun.
\end{lemma}

\noindent {\bf Proof of Lemma~\ref{lem:sunflower}.} Consider all the
sunflowers of $G$ satisfying the conditions of the lemma that have
the fewest number of vertices. From these, pick the sunflower
$S=((P_1,P_2, \ldots, P_k),$ $( c_1, c_2, \ldots, c_k))$ so that $k$
is as small as possible. We may assume that $P_k = {c_k}x{c_1}$ and
$c_1$ sees $c_k$. Also, recall that the two endpoints of $P_i$, $1
\leq i \leq k-1$, are $c_i$ and $c_{i+1}$.

Suppose $k = 3$. Then, it is seen that $S$ is indeed a 3-sun (or
else there is a building in $G$). Therefore, we can assume that $k
\geq 4$.

First, we show that every center vertex must see $c_k$. It will then
follow from symmetry that every center vertex must see $c_1$ also.

Consider a $c_i$ such that $i \notin \{1, k\}$. Suppose $c_i$ misses
$c_k$ ($i = 2$, $i = k-1$ are possible). If $c_i$ misses every
center vertex of $S$, then $((P_1, \ldots ,P_{i-1}P_i, \ldots,
P_k),$ $( c_1, \ldots, c_{i-1}, c_{i+1}, \ldots, c_k))$ is a
sunflower on the same set of vertices as $S$ that contains fewer
center vertices and in which $x$ is a tip, a contradiction.
Therefore, we can assume that $c_i$ misses $c_k$, but sees some
center vertex of $S$.

Suppose $c_i$ misses every $c_j$, $i < j < k$. Traversing in the
clockwise direction from $c_1$, let $c_t$ be the first center vertex
that $c_i$ sees. Suppose $t = 1$. If $i = k-1$, then $\{x, c_1\}
\cup P_{k-1}$ induces a building. Therefore, $i \not= k-1$. Then,
$((c_1P_i, P_{i+1}, \ldots, P_k),$ $( c_1, c_{i+1}, \ldots, c_k))$
is a smaller sunflower in which $x$ is a tip. Now suppose $1 < t <
i$. Then, $((P_1, \ldots, P_{t-1}, c_tP_i, P_{i+1}, \ldots, P_k),
(c_1, \ldots,\\ c_{t-1}, c_t, c_{i+1}, \ldots, c_k))$ is a smaller
sunflower in which $x$ is a tip.

Now we can assume that $c_i$ sees some $c_j$ such that $i < j < k$
(and therefore, $i \not= k-1$). Let $c_t$ be the first center vertex
in the counter clockwise direction from $c_k$ that $c_i$ sees such
that $i < t < k$. If $c_i$ has no neighbour $c_j$ such that $1 \leq
j < i$, then $((P_1, \ldots, P_{i-1}c_t, P_t, \ldots, P_k), (c_1,
\ldots, c_{i-1}, c_t, \ldots, c_k))$ is a smaller sunflower in which
$x$ is a tip. Otherwise, let $c_r$ be the first center vertex in the
clockwise direction from $c_1$ that $c_i$ sees such that $1 \leq r <
i$. If $r = 1$, then $(({c_1}{c_i}{c_t}, P_t, \ldots, P_k), (c_1,
c_t, \ldots, c_k))$ is a smaller sunflower in which $x$ is a tip. On
the other hand, if $r \not= 1$, then $((P_1, \ldots, P_{r-1},
{c_r}{c_i}{c_t}, P_t, \ldots, P_k)$, $(c_1, \ldots, c_{r-1}, c_r,
c_t, \ldots, c_k))$ is a smaller sunflower in which $x$ is a tip.

We can now conclude that every center vertex sees $c_k$ and every
center vertex sees $c_1$ also.

Next, we want to show that for $1 \leq i \leq k-1$, $c_i$ sees
$c_{i+1}$ and every $P_i$ must have exactly two edges. Whenever such
a $P_i$ is established to have exactly two edges, $d_i$ will refer
to the only petal vertex of $P_i$.

We already have established that $c_1$ sees $c_2$ and $c_{k-1}$ sees
$c_k$. Note that for any $i$, if $c_i$ sees $c_{i+1}$ then $P_i$ has
length two; for otherwise, $P_i \cup \{c_k\}$ or $P_i \cup \{c_1\}$
induces a building. Let $i$ be the smallest subscript (if it exists)
such that $c_i$ misses $c_{i+1}$. Now, if $c_{i-1}$ sees $c_{i+1}$,
then $P_i \cup \{c_{i-1}, d_{i-1} \}$ induces a building; otherwise,
$P_i \cup \{ c_k, c_{i-1} \}$ induces a building.

Finally, it can now be seen that any $d_i$ can play the role of
vertex $x$. Therefore, applying the same arguments to every $d_i$ we
can conclude that the set of center vertices forms a clique and $S$
is indeed a sun. \hfill $\qed$

We note that an efficient algorithm to find a sun in a building-free
graph, when it exists, can be extracted from the proofs of
Lemma~\ref{lem:flower} and Lemma~\ref{lem:sunflower}.

We now consider Algorithm {\it in-3-sun}, which is called by
Algorithm {\it tip-of-sunflower}. The algorithm {\it in-3-sun} is
reproduced below from \cite{EHS} for the sake of completeness.
However, we refer the reader to \cite{EHS} for the details on its
correctness and complexity.

\bt xx\=xxxx\=xxxx\=xxxx\=xxxx\=xxxx\kill
{\bf Algorithm} {\it in-3-sun}\\
{\bf Input}:  Chordal graph $H$, simplicial vertex $x$ of $H$, and
\\
\>\> ordering $\ord$ $= {y_1},{y_2}, \ldots, {y_k}$ of $M(x)$ such that\\
\> \>  $\ord$ is a strong elimination ordering for
$H[M(x)]$, \\
\> \> for $i < j$, $|N({y_i}) \cap N(x)| \leq |N({y_j}) \cap
N(x)|$, and \\
 \> \> for $i < j$ with ${y_i}{y_j} \in E(H)$,
$(N({y_i}) \cap N(x)) \subseteq (N({y_j}) \cap
N(x))$. \\
{\bf Output}: {\it true} if $x$ is in a 3-sun in $H$; {\it false} otherwise\\
\\
\{\\
\>1.\> {\bf for} $i$ = 1 to $k-2$ {\bf do}\\
\> \> \> For $i < p < q$ such that ${y_i}{y_q}{y_p}$ is a $P_3$ in $H$, \\
\> \> \> \> mark $[{y_i}, {y_p}]$ as a red edge to be added. \\
\> \> {\bf end} \\
\>2.\> Add all the red edges to $H$ to obtain $H'$. \\
\>3.\> Let ${x_1},{x_2},\ldots,{x_r}$ be an arbitrary ordering of $N(x)$. \\
\> \> {\bf if} ${y_1},\ldots,{y_k},{x_1},\ldots,{x_r},x$ is a perfect elimination scheme for $H'$ {\bf then}\\
\> \> \>         return ({\it false})\\
\> \> {\bf else}\\
\> \> \>         return ({\it true});\\
\} \et

\noindent
\begin{lemma} \label{lem:rededge} \normalfont{\cite{EHS}}
Let $G$ be a strongly chordal graph with the strong elimination
ordering ${\cal S} = {v_1},{v_2},\ldots,{v_n}$. Let $G^\prime$ be
the graph obtained from $G$ by adding every edge ${v_i}{v_j}$ such
that, for some $k$ with $i < j < k$, ${v_i}{v_k}{v_j}$ is a $P_3$ in
$G$. Then, $G^\prime$ is a chordal graph and ${\cal S}$ is a perfect
elimination scheme for $G^\prime$.
\end{lemma}

\begin{lemma} \label{lem:reddom} \normalfont{\cite{EHS}}
Vertex $x$ is in a 3-sun in $H$ if and only if for some red edge
${y_i}{y_j}$ added with $i < j$, $y_j$ does not $N(x)$-dominate
$y_i$.
\end{lemma}

\noindent
\begin{lemma} \label{lem:3-sun-corr} \normalfont{\cite{EHS}}
Vertex $x$ is in a 3-sun in $H$ if and only if ${\cal S} =
{y_1},\ldots,{y_k},{x_1},\ldots,{x_r},x$ is not a perfect
elimination scheme for $H^\prime$.
\end{lemma}

\noindent
\begin{corollary} \label{cor:3-sun-corr} \normalfont{\cite{EHS}}
${\cal S} = {y_1},\ldots,{y_k},{x_1},\ldots{x_r},x$ is not a perfect
elimination scheme for $H'$ if and only if there exists  $y_i \in
M(x)$ such that $y_j \in M(x)$ is the first neighbour of $y_i$ in
$H'$ that comes after it in $\ord$ and there exists $w \in N(x)$
that $y_i$ sees but $y_j$ misses.
\end{corollary}

\subsection{Time complexity}

\begin{lemma} \label{lem:completion} \normalfont{\cite{EHS}}
Let $G$ be a graph and $\ord = {v_1},{v_2},\ldots, {v_n}$ be an
arbitrary ordering of its vertices. Then, strongly chordal
completion of $G$ ensuring that $\ord$ is a strong elimination
ordering for the resulting graph can be performed in O($n^2$) time.
\end{lemma}

\noindent
\begin{lemma} \label{lem:3-sun-comp} \normalfont{\cite{EHS}}
Algorithm {\it in-3-sun} can be implemented to run in O($n^2$) time.
\end{lemma}

\begin{lemma} \label{lem:flower-compl} \normalfont{\cite{EHS}}
Algorithm {\it tip-of-sunflower} can be implemented to run in
O($n^2$) time.
\end{lemma}
\noindent {\bf Proof of Lemma~\ref{lem:flower-compl}.} Steps 1 and 2
can easily be done in linear time. By Lemma~\ref{lem:completion},
step~3 can be done in O($n^2$) time. Finally, by
Lemma~\ref{lem:3-sun-comp}, step~4 can also be done in O($n^2$)
time. \hfill $\qed$

\begin{theorem}\label{thm:suncompl}
Algorithm {\it find-sun} is correct and it runs in
\normalfont{O(min}$\{m{n^3}, m^{1.5}n^2\})$ time.
\end{theorem}

\noindent {\bf Proof of Theorem~\ref{thm:suncompl}.}
It is clear from Lemma~\ref{lem:sun}, Lemma~\ref{lem:flower}, and
Lemma~\ref{lem:sunflower} that $G'$ contains a sun if and only if
for some vertex $x$ and some pair of adjacent neighbours $y$ and $z$
of $x$, $x$ is a tip of a sunflower in the graph $G$ constructed by
the algorithm. Therefore, the algorithm is correct.

The algorithm {\it find-sun} invokes algorithm {\it
tip-of-sunflower} O($mn$) times. Since each such invocation costs
O($n^2$) time, the overall time complexity of the algorithm is
O($m{n^3}$).

Since all the triangles of a graph can be listed in O($m^{1.5}$)
time \cite{AYZ}, finding a building as well as algorithm {\it
find-sun} can be implemented to run in O($m^{1.5}{n^2}$) time also.
\hfill $\qed$

\section{A class of (building, sun)-free graphs}
Several classes of graphs have been characterized via the presence
of an elimination scheme of vertices \cite{B}. There are also graph
classes that are known to admit an interesting elimination scheme of
vertices, though the presence of such a scheme may not characterize
the class of graphs.

A {\it bull} is the graph constructed by starting with a $P_4$
$abcd$ and then adding vertex $e$ that is adjacent only to the
vertices $b$ and $c$. The vertex $e$ is said to be the {\it nose} of
the bull.

In this section we consider the class of (building, gem)-free
graphs. It is seen that every distance hereditary graph ((house,
hole, domino, gem)-free graph) is (building, gem)-free and that
every (building, gem)-free graph is (building, sun)-free. We show
that every (building, gem)-free graph admits an elimination scheme
of vertices that generalizes a known elimination scheme of vertices
\cite{DN} for distance hereditary graphs. We show that, as in the
case of distance hereditary graphs, a lexicographic breadth first
search (LBFS) can be used to generate the elimination scheme. Next,
we reproduce the details of LBFS from \cite{G}. It is well known
that LBFS on any graph can be implemented to run in linear time
\cite{RTL}.

\bt xx\=xxxx\=xxxx\=xxxx\=xxxx\=xxxx\kill
{\bf Algorithm} {\it LBFS}\\
{\bf Input}:  graph $G$ \\
{\bf Output}: ordering $\sigma = {v_1}{v_2} \ldots {v_n}$ of vertices of $G$\\
\\
\{\\
\>Assign the label $\emptyset$ to each vertex. \\
\> {\bf for} $i$ = $n$ {\bf downto} 1 {\bf do} \\
\> \{ \\
\> \> Choose an unnumbered vertex $w$ with the largest label. \\
\> \> $v_i = w$ \\
\> \> {\bf for} each unnumbered neighbour $x$ of $w$ {\bf do} \\
\> \> \> Append $i$ to the label of $x$. \\
\> \}\\
\}
\et

We refer to the output $\sigma$ of LBFS on graph $G$ as an {\it LBFS
ordering of G}.  For such an ordering $\sigma$, we say ``$v_i <
v_j$'' and ``$v_j$ is to the right of $v_i$" to mean vertex $v_i$
precedes vertex $v_j$ in $\sigma$. The following well-known property
\cite{RTL} of an LBFS ordering of a graph $G = (V, E)$ will be used
repeatedly:

\begin{enumerate}
\item[$(P_*)$] For vertices $a, b, c \in V$, if $a < b < c$, $ac \in E$, and
$bc \notin E$, then there exists $d \in V$ such that $c < d$, $bd
\in E$, and $ad \notin E$.
\end{enumerate}

We refer to an ordered triple $(a, b, c)$ of vertices that satisfies
the conditions of $(P_*)$ as an {\it extendable triple}.

\begin{figure}[h]
\epsfxsize=8cm \centerline{\epsfbox{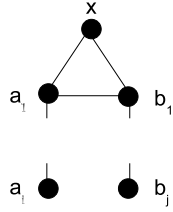}}
%
%
\vskip -4cm
\caption{An extended bull in which $x$ is the nose. We have $i \geq
2$, $j \geq 2$, and each of ${a_1} \ldots {a_i}$ and ${b_1} \ldots
{b_j}$ is an induced path.} \label{fig:extbull}
\end{figure}

\begin{definition} \label{def:extbull}
An extended bull is the graph obtained starting from an induced path
${a_i} \ldots {a_1}{b_1} \ldots {b_j}$, where $i \geq 2$ and $j \geq
2$, and adding vertex $x$ that sees only $a_1$ and $b_1$. Therefore,
when $i = 2$ and $j = 2$, an extended bull is a bull. Vertex $x$ is
the nose of the extended bull. We refer to such an extended bull
(bull) as $B(x, a_i, b_j)$ ($B(x, a_2, b_2)$).
\end{definition}

\begin{definition} \label{def:rnose}
Suppose $\sigma$ is an ordering of vertices of graph $G$ produced by
LBFS. For a vertex $x$, let $H_x$ be the subgraph of $G$ induced by
$x$ and the vertices to the right of $x$ in $\sigma$. By ``$B(x,
a_2, b_2)$ is the rightmost bull in which x is the nose'' we mean
$B(x, a_2, b_2)$ is a bull in $H_x$ and no vertex to the right of a
member of $\{a_2, a_1, b_1, b_2 \}$ in $\sigma$ and $x$ are in a
bull in $H_x$ in which $x$ is the nose.
\end{definition}

\begin{definition} \label{def:nearbuilding}
A near building is a graph $G=(\{v, w_1, \ldots, w_j\} (j \geq 4),
E)$ such that $\{vw_1, vw_j, w_1w_j\} \subset E$, $w_iw_{i+1} \in E$
for all $1 \leq i \leq j-1$, and the only other edges in $E$, if
they exist, are of the form $w_jw_i$ where $2 \leq i \leq j-2$.
\end{definition}

\begin{claim} \label{claim:nearbuilding}
Let $G$ be a (building, gem)-free graph. Then $G$ cannot contain a
near building as an induced subgraph.
\end{claim}
{\bf Proof of Claim~\ref{claim:nearbuilding}.} Suppose $G$ contains
an induced near building.  If $w_j$ misses $w_2$, then $\{v, w_j,
w_1, \ldots, w_i\}$, where $w_i$ is the smallest indexed vertex that
$w_j$ sees after $w_1$, induces a building in $G$. If $w_j$ sees
$w_2$ and $w_3$, then $G$ contains an induced gem. If $w_j$ sees
$w_2$ and misses $w_3$, then $\{w_1, w_j, w_2, \ldots, w_i\}$, where
$w_i$ is the smallest indexed vertex that $w_j$ sees after $w_2$,
induces a building in $G$. In each case a contradiction is reached;
therefore, $G$ cannot contain an induced near building. \hfill \qed

We need the following lemma.

\begin{lemma} \label{lem:extend}
Suppose $G$ is a (building, gem)-free graph and $\sigma$ is an
ordering of vertices of $G$ produced by LBFS. Suppose vertex $x$ is
the nose of a bull in $H_x$. Further, suppose the following hold:
\begin{itemize}
 \item $B(x, a_2, b_2)$ is the rightmost bull in which $x$ is the
 nose.
 \item Vertex $x$ is the nose of the extended bull $B(x, a_i, b_j)$ in
 $H_x$ where $i \geq 2$ and $j \geq 2$.
 \item For all $r$, $2 \leq r \leq i-1$, $a_r < a_{r+1}$ in $\sigma$.
 \item For all $r$, $2 \leq r \leq j-1$, $b_r < b_{r+1}$ in $\sigma$.
 \item When $i+1 = 3$, at least one of the following holds:  $a_{i+1}$ misses $x$,
$a_{i+1}$ misses $b_1$, or $a_1 < a_{i+1}$.
 \item When $j+1 = 3$, at least one of the following holds:  $b_{j+1}$ misses $x$,
$b_{j+1}$ misses $a_1$, or $b_1 < b_{j+1}$.
\item For every $k$, $4 \leq k \leq i$, $a_k$ is the rightmost vertex in $\sigma$
that sees $a_{k-1}$ and misses every $b_r$, $1 \leq r \leq j$.
\item For every $k$, $4 \leq k \leq j$, $b_k$ is the rightmost vertex in $\sigma$
that sees $b_{k-1}$ and misses every $a_r$, $1 \leq r \leq i$.
\item Either $a_3$ is the rightmost
vertex in $\sigma$ that sees $a_2$ and misses $x$ or $a_3$ is the
rightmost vertex in $\sigma$ that sees $a_2$ and misses every $b_r$,
$1 \leq r \leq j$.
\item Either $b_3$ is the rightmost
vertex in $\sigma$ that sees $b_2$ and misses $x$ or $b_3$ is the
rightmost vertex in $\sigma$ that sees $b_2$ and misses every $a_r$,
$1 \leq r \leq i$.
\end{itemize}
Then we have the following:
\begin{enumerate}
\item If $i+1 = 3$, $x < a_2 < b_1$ in $\sigma$, and $a_{i+1}$ is the
rightmost vertex after $b_1$ in $\sigma$ such that $a_{i+1}$ sees
$a_i$ but misses $x$, then $B(x, a_{i+1}, b_j)$ is a larger extended
bull in which $x$ is the nose.
\item If $b_{j-1} < a_i < b_j$ in $\sigma$ and $a_{i+1}$ is the
rightmost vertex after $b_j$ in $\sigma$ such that $a_{i+1}$ sees
$a_i$ but misses $b_{j-1}$, then $B(x, a_{i+1}, b_j)$ is a larger
extended bull in which $x$ is the nose.
\item If $j+1 = 3$, $x < b_2 < a_1$ in $\sigma$, and $b_{j+1}$ is the
rightmost vertex after $a_1$ in $\sigma$ such that $b_{j+1}$ sees
$b_j$ but misses $x$, then $B(x, a_i, b_{j+1})$ is a larger extended
bull in which $x$ is the nose.
\item If $a_{i-1} < b_j < a_i$ in $\sigma$ and $b_{j+1}$ is the
rightmost vertex after $a_i$ in $\sigma$ such that $b_{j+1}$ sees
$b_j$ but misses $a_{i-1}$, then $B(x, a_{i}, b_{j+1})$ is a larger
extended bull in which $x$ is the nose.
\end{enumerate}
\end{lemma}

{\bf Proof of Lemma~\ref{lem:extend}.} Let $G$ be a (building,
gem)-free graph and $\sigma$ be an ordering of vertices of $G$
produced by LBFS. Suppose vertex $x$ is the nose of an extended bull
in $H_x$ that satisfies the hypotheses of the lemma. We will prove
Statements~1 and 2 simultaneously; the proofs of Statements~3 and 4
are similar. In the case of Statement~1, suppose $x < a_2 < b_1$ in
$\sigma$ and $a_3$ is the rightmost vertex after $b_1$ in $\sigma$
such that $a_3$ sees $a_2$ but misses $x$.  In the case of
Statement~2, suppose $b_{j-1} < a_i < b_j$ in $\sigma$ and $a_{i+1}$
is the rightmost vertex after $b_j$ in $\sigma$ such that $a_{i+1}$
sees $a_i$ but misses $b_{j-1}$.  We will prove that $B(x, a_{i+1},
b_j)$ is a larger extended bull in which $x$ is the nose.

We need to establish that the new vertex $a_{i+1}$ misses $x$. This
follows from the hypothesis in the case of Statement~1.  For
Statement~2 we have the following argument.  Suppose $a_{i+1}$ sees
$x$ and $b_1$. From the hypotheses we have the following: If
$i+1=3$, then $a_1 < a_{i+1}$; otherwise $a_i < a_{i+1}$. Now, if
$a_{i+1}$ misses $b_2$, then $B(x, a_2, b_2)$ is not the rightmost
bull in which $x$ is the nose; $\{x, a_{i+1}, a_i, b_1, b_2\}$
induces a bull that has a vertex to the right of a vertex of $B(x,
a_2, b_2)$. So $a_{i+1}$ sees $b_2$. Now, if $a_{i+1}$ misses $a_1$,
$\{b_1, b_2, a_{i+1}, x, a_1\}$ induces a gem in $G$; hence,
$a_{i+1}$ must see $a_1$. If $i+1 = 3$, then $\{a_3, a_2, a_1, b_1,
b_2\}$ induces a gem in $G$. So assume that $i+1 \geq 4$. Then $\{x,
a_1, \ldots, a_{i+1}\}$ induces a near building in $G$, which by
Claim~\ref{claim:nearbuilding} is a contradiction. So now suppose
$a_{i+1}$ sees $x$ and misses $b_1$. If $a_{i+1}$ misses $a_1$,
consider $r$, $2 \leq r \leq i$, the smallest index such that
$a_{i+1}$ sees $a_r$.  Then $\{x, b_1, a_{i+1}, a_1, \ldots, a_r\}$
induces a building in $G$.   So $a_{i+1}$ sees $a_1$.  If $i+1=3$,
$\{a_1, a_2, a_3, x, b_1\}$ induces a gem in $G$.  If $i+1 \geq 4$,
then $\{x, a_1, \ldots, a_{i+1}\}$ induces a near building in $G$.
So we may conclude that $a_{i+1}$ misses $x$.

The remainder of the argument applies to Statements~1 and 2. Suppose
$a_{i+1}$ sees $a_1$. Further suppose $a_{i+1}$ misses $b_1$. If
$a_{i+1}$ sees $b_2$, $G$ has an induced building. If $a_{i+1}$
misses $b_2$, then $B(x, a_2, b_2)$ is not the rightmost bull in
which $x$ is the nose; $\{x, a_1, a_{i+1}, b_1, b_2\}$ induces a
bull that has a vertex to the right of a vertex of $B(x, a_2, b_2)$.
Therefore, $a_{i+1}$ must see $b_1$. If $i+1 = 3$, then $\{a_1, a_2,
a_3, b_1, x\}$ induces a gem in $G$. So assume that $i+1 \geq 4$.
Then $\{b_1, a_1, \ldots, a_{i+1}\}$ induces a near building in $G$.
So we may conclude that $a_{i+1}$ misses $a_1$.

Now we claim that $a_{i+1}$ misses all of $\{a_1, \ldots,
a_{i-1}\}$. This is clearly the case if $i+1 = 3$, so assume that
$i+1 \geq 4$. We have already established that $a_{i+1}$ misses $x$
and $a_1$. Suppose $a_{i+1}$ sees some vertex in $\{a_2, \ldots,
a_{i-1}\}$. Let $r$, $2 \leq r \leq i-1$, be the smallest index such
that $a_{i+1}$ sees $a_r$. Suppose $r=2$ and $a_3$ is the rightmost
vertex that sees $a_2$ and misses $x$.  In this case, we have
$a_{i+1}$ is a vertex to the right of $a_3$ that sees $a_2$ and
misses $x$, which is a contradiction. Otherwise, $r \geq 3$ or $a_3$
is the rightmost vertex that sees $a_2$ and misses every $b_r$, $1
\leq r \leq j$.  In these cases, by the choice of $a_{r+1}$,
$a_{i+1}$ sees some vertex in $\{b_1, \ldots, b_j\}$; otherwise,
$a_{i+1}$ is a vertex to the right of $a_{r+1}$ that sees $a_r$ and
misses all of $\{b_1, \ldots, b_j\}$. Let $s$, $1 \leq s \leq j$, be
the smallest index such that $a_{i+1}$ sees $b_s$.  Now $\{x, a_1,
\ldots, a_r, a_{i+1}, b_1, \ldots, b_s\}$ induces a building in $G$,
which is a contradiction.

Finally, $a_{i+1}$ misses all of $\{b_1, \ldots, b_j\}$.  Suppose
not. Let $s$, $1 \leq s \leq j$, be the smallest index such that
$a_{i+1}$ sees $b_s$.  Now $\{x, a_1, \ldots, a_{i+1}, b_1, \ldots,
b_s\}$ induces a building in $G$, which is a contradiction.

Hence, $B(x, a_{i+1}, b_j)$ is a larger extended bull in which $x$
is the nose. \hfill \qed

Recall that if $\sigma$ is an LBFS ordering of graph $G$, then $H_x$
refers to the subgraph of $G$ induced by $x$ and the vertices to the
right of $x$ in $\sigma$.

\begin{theorem} \label{thm:scheme} Suppose $G$ is a (building,
gem)-free graph and $\sigma$ is an LBFS ordering of $G$. Then, each
vertex $v$ is not the nose of any bull in $H_v$.
\end{theorem}
{\bf Proof of Theorem~\ref{thm:scheme}.} Let us suppose to the
contrary that some vertex $x$ is the nose of a bull in $H_x$. Let
$B(x, a_2, b_2)$ be the rightmost bull in $H_x$ in which $x$ is the
nose. We will show that starting from $B(x, a_2, b_2)$, we can
inductively grow larger and larger extended bulls in $H_x$ in which
$x$ is the nose, contradicting the graph $G$ being finite. From here
on, all the adjacencies referred to are with respect to the graph
$H_x$.

Without loss of generality, we can assume $x < a_1 < b_1$. First, we
consider the twelve possible arrangements of the vertices $x, a_1 ,
a_2 , b_1$, and $b_2$ in $\sigma$ and for each arrangement we show
that $H_x$ contains a specific larger extended bull $B(x, a_i,
b_j)$. We do this so that every extended bull $B(x, a_i, b_j)$ that
we develop is then conducive to the same inductive argument. Having
done this basis step, we will use the inductive argument to show
that any extended bull $B(x, a_i, b_j)$ can further be extended into
either $B(x, a_{i+1}, b_j)$ or $B(x, a_i, b_{j+1})$.

We repeatedly use $(P_*)$ on an extendable triple $(a, b, c)$ that
is part of some extended bull $L$ to bring in a new vertex $d$ into
the argument; for the sake of brevity, we say ``use the triple $(a,
b, c)$'' to refer to this process. When we do this, we will invoke
Lemma~\ref{lem:extend} to conclude that the new vertex $d$ and the
vertices of $L$ induce a larger extended bull. Also, after each
extension, we designate a specific extendable triple $(x, y, z)$ as
the ``next triple''. The idea is that in the next step of extension,
the triple $(x, y, z)$ is to be used. We carefully choose such a
next triple so as to ensure that the conditions of
Lemma~\ref{lem:extend} are met. Finally, every time we bring in a
new vertex $d$ while extending a triple $(a, b, c)$, we always
choose the rightmost such $d$ in $\sigma$ that satisfies the
conditions of selection. We now consider each of the twelve
possibilities.

\noindent
{\bf Case 1:} $x\ \ a_1\  \ b_1\  \ a_2\  \ b_2$\\
Use the triple $(b_1, a_2, b_2)$ and choose $a_3 > b_2$ that sees
$a_2$ and misses $b_1$. Apply Lemma~\ref{lem:extend} to get the
larger extended bull $B(x, a_3, b_2)$ and the sequence $x\ \ a_1\  \
b_1\ \ a_2\  \ b_2\ \ a_3$ of vertices.  The next triple is $(a_2,
b_2, a_3)$. Note that if this triple is used to choose $b_3 > a_3$
that sees $b_2$ and misses $a_2$, then $b_1 < b_3$ will hold.

\noindent
{\bf Case 2:} $x\ \ a_1\ \ b_1\ \ b_2\ \ a_2$\\
Use the triple $(a_1, b_2, a_2)$ and choose $b_3 > a_2$ that sees
$b_2$ and misses $a_1$. Apply Lemma~\ref{lem:extend} to get the
extended bull $B(x, a_2, b_3)$ and the sequence $x\ \ a_1\ \ b_1\ \
b_2\ \ a_2\ \ b_3$ of vertices. The next triple is $(b_2, a_2,
b_3)$. Note that if this triple is used to choose $a_3 > b_3$ that
sees $a_2$ and misses $b_2$, then $a_1 < a_3$ will hold.

\noindent
{\bf Case 3:} $x\ \ a_1\ \ a_2\ \ b_1\ \ b_2$\\
Use the triple $(x, a_2, b_1)$ to choose $a_3 > b_1$ that sees $a_2$
and misses $x$. By Lemma~\ref{lem:extend}, $B(x, a_3, b_2)$ is a
larger extended bull. We divide into two subcases based on the
resulting sequence of vertices.

\noindent {\bf Case 3.1:} Suppose the resulting sequence of vertices
is $x\ \ a_1\ \ a_2\ \ b_1\ \ a_3\ \ b_2$. Use the triple  $(b_1,
a_3, b_2)$ and choose $a_4 > b_2$ that sees $a_3$ and misses $b_1$.
Apply Lemma~\ref{lem:extend} to get the extended bull $B(x, a_4,
b_2)$. The resulting sequence is $x\ \ a_1\ \ a_2\ \ b_1\ \ a_3\ \
b_2\ \ a_4$ and the next triple is $(a_3, b_2, a_4)$. Note that if
this triple is used to choose $b_3 > a_4$ that sees $b_2$ and misses
$a_3$, then $b_1 < b_3$ will hold.

\noindent {\bf Case 3.2:} Suppose the sequence of vertices is $x\ \
a_1\ \ a_2\ \ b_1\ \ b_2\ \ a_3$. Then, the next triple is $(a_2,
b_2, a_3)$. Note that if this triple is used to choose $b_3 > a_3$
that sees $b_2$ and misses $a_2$, then $b_1 < b_3$ will hold.

\noindent
{\bf Case 4:} $x\ \ a_1\ \ b_2\ \ b_1\ \ a_2$\\
Use the triple $(a_1, b_2, a_2)$ and choose $b_3 > a_2$ that sees
$b_2$ and misses $a_1$. Apply Lemma~\ref{lem:extend} to get the
extended bull $B(x, a_2, b_3)$ and the sequence $x\ \ a_1\ \ b_2\ \
b_1\ \ a_2\ \ b_3$ of vertices. The next triple is $(b_2, a_2,
b_3)$. Note that if this triple is used to choose $a_3 > b_3$ that
sees $a_2$ and misses $b_2$, then $a_1 < a_3$ will hold.

\noindent
{\bf Case 5:} $x\ \ a_1\ \ a_2\ \ b_2\ \ b_1$\\
Use the triple $(x, a_2, b_1)$ and choose $a_3 > b_1$ that sees
$a_2$ and misses $x$. Apply Lemma~\ref{lem:extend} to get the
extended bull $B(x, a_3, b_2)$ and the sequence $x\ \ a_1\ \ a_2\ \
b_2\ \ b_1\ \ a_3$ of vertices. The next triple is $(a_2, b_2,
a_3)$. Note that if this triple is used to choose $b_3 > a_3$ that
sees $b_2$ and misses $a_2$, then $b_1 < b_3$ will hold.

\noindent
{\bf Case 6:} $x\ \ a_1\ \ b_2\ \ a_2\ \ b_1$\\
Use the triple $(a_1, b_2, a_2)$ and choose $b_3 > a_2$ that sees
$b_2$ and misses $a_1$. Apply Lemma~\ref{lem:extend} to get the
extended bull $B(x, a_2, b_3)$. We subdivide this case based on the
resulting sequence of vertices.

\noindent {\bf Case 6.1:} Suppose the resulting sequence of vertices
is $x\ \ a_1\ \ b_2\ \ a_2\ \ b_3\ \  b_1$. Use the triple $(x, a_2,
b_1)$ and choose $a_3 > b_1$ that sees $a_2$ and misses $x$. Apply
Lemma~\ref{lem:extend} to get the extended bull $B(x, a_3, b_3)$.
The resulting sequence is $x\ \ a_1\ \ b_2\ \ a_2\ \ b_3\ \  b_1\ \
a_3$ and the next triple is $(a_2, b_3, a_3)$.

\noindent {\bf Case 6.2:} Suppose the resulting sequence of vertices
is $x\ \ a_1\ \ b_2\ \ a_2\ \ b_1\ \ b_3$. Use the triple $(x, a_2,
b_1)$ and choose $a_3 > b_1$ that sees $a_2$ and misses $x$. Apply
Lemma~\ref{lem:extend} to get the extended bull $B(x, a_3, b_3)$. If
the resulting sequence is $x\ \ a_1\ \ b_2\ \ a_2\ \ b_1\ \ a_3\ \
b_3$, the next triple is $(b_2, a_3, b_3)$. Otherwise, the resulting
sequence is $x\ \ a_1\ \ b_2\ \ a_2\ \ b_1\ \ b_3\ \ a_3$ and the
next triple is $(a_2, b_3, a_3)$.

\noindent
{\bf Case 7:} $x\ \ a_2\ \ a_1\ \ b_1\ \ b_2$\\
Use the triple $(x, a_2, b_1)$ to choose $a_3 > b_1$ that sees $a_2$
and misses $x$. Apply Lemma~\ref{lem:extend} to get the extended
bull $B(x, a_3, b_2)$. We subdivide based on the resulting sequence.

\noindent {\bf Case 7.1:} Suppose the resulting sequence is $x\ \
a_2\ \ a_1\ \ b_1\ \ a_3\ \ b_2$. Use the triple $(b_1, a_3, b_2)$
and choose $a_4 > b_2$ that sees $a_3$ and misses $b_1$. Apply
Lemma~\ref{lem:extend} to get the extended bull $B(x, a_4, b_2)$ and
the sequence $x\ \ a_2\ \ a_1\ \ b_1\ \ a_3\ \ b_2\ \ a_4$. The next
triple is $(a_3, b_2, a_4)$. Note that if this triple is used to
choose $b_3 > a_4$ that sees $b_2$ and misses $a_3$, then $b_1 <
b_3$ will hold.

\noindent {\bf Case 7.2:} Suppose the resulting sequence is $x\ \
a_2\ \ a_1\ \ b_1\ \ b_2\ \ a_3$. Then, the next triple is $(a_2,
b_2, a_3)$. Note that if this triple is used to choose $b_3 > a_3$
that sees $b_2$ and misses $a_2$, then $b_1 < b_3$ will hold.

\noindent
{\bf Case 8:} $x\ \ a_2\ \ a_1\ \ b_2\ \ b_1$\\
Use the triple $(x, a_2, b_1)$ and choose $a_3 > b_1$ that sees
$a_2$ and misses $x$. Apply Lemma~\ref{lem:extend} to get the larger
extended bull $B(x, a_3, b_2)$. The resulting sequence is $x\ \ a_2\
\ a_1\ \ b_2\ \ b_1\ \ a_3$ and the next triple is $(a_2, b_2,
a_3)$. Note that if this triple is used to choose $b_3 > a_3$ that
sees $b_2$ and misses $a_2$, then $b_1 < b_3$ will hold.

\noindent
{\bf Case 9:} $x\ \ b_2\ \ a_1\ \ b_1\ \ a_2$\\
Use the triple $(x, b_2, a_1)$ and choose $b_3 > a_1$ that sees
$b_2$ and misses $x$. Apply Lemma~\ref{lem:extend} to get the larger
extended bull $B(x, a_2, b_3)$. There are three possibilities for
the resulting sequence.

\noindent {\bf Case 9.1:} Suppose the resulting sequence is $x\ \
b_2\ \ a_1\ \ b_3\ \ b_1\ \ a_2$. Use the triple $(a_1, b_3, a_2)$
and choose $b_4 > a_2$ that sees $b_3$ that misses $a_1$. Apply
Lemma~\ref{lem:extend} to get the extended bull $B(x, a_2, b_4)$ and
the sequence $x\ \ b_2\ \ a_1\ \ b_3\ \ b_1\ \ a_2\ \ b_4$. The next
triple is $(b_3, a_2, b_4)$. Note that if this triple is used to
choose $a_3 > b_4$ that sees $a_2$ and misses $b_3$, then $a_1 <
a_3$ will hold.

\noindent {\bf Case 9.2:} Suppose the resulting sequence is $x\ \
b_2\ \ a_1\ \ b_1\ \ b_3\ \ a_2$. Use the triple $(a_1, b_3, a_2)$
and choose $b_4 > a_2$ that sees $b_3$ that misses $a_1$. Apply
Lemma~\ref{lem:extend} to get the extended bull $B(x, a_2, b_4)$ and
the sequence $x\ \ b_2\ \ a_1\ \ b_1\ \ b_3\ \ a_2\ \ b_4$. The next
triple is $(b_3, a_2, b_4)$. Note that if this triple is used to
choose $a_3 > b_4$ that sees $a_2$ and misses $b_3$, then $a_1 <
a_3$ will hold.

\noindent {\bf Case 9.3:} Suppose the resulting sequence is $x\ \
b_2\ \ a_1\ \ b_1\ \ a_2\ \ b_3$. The next triple is $(b_2, a_2,
b_3)$. Note that if this triple is used to choose $a_3 > b_3$ that
sees $a_2$ and misses $b_2$, then $a_1 < a_3$ will hold.

\noindent
{\bf Case 10:} $x\ \ b_2\ \ a_1\ \ a_2\ \ b_1$\\
Use the triple $(x, b_2, a_1)$ to choose $b_3 > a_1$ that sees $b_2$
and misses $x$. Use Lemma~\ref{lem:extend} to get the larger
extended bull $B(x, a_2, b_3)$. Again, there are three possible
resulting sequences.

\noindent {\bf Case 10.1:} Suppose the resulting sequence is $x\ \
b_2\ \ a_1\ \ b_3\ \ a_2\ \ b_1$. Use the triple is $(a_1, b_3,
a_2)$ and choose $b_4 > a_2$ that sees $b_3$ and misses $a_1$. Apply
Lemma~\ref{lem:extend} to get the extended bull $B(x, a_2, b_4)$.
The resulting sequence is $x\ \ b_2\ \ a_1\ \ b_3\ \ a_2\ \ b_4\ \
b_1$ or $x\ \ b_2\ \ a_1\ \ b_3\ \ a_2\ \ b_1\ \ b_4$. In each case,
the next triple is $(b_3, a_2, b_4)$. Note that if this triple is
used to choose $a_3 > b_4$ that sees $a_2$ and misses $b_3$, then
$a_1 < a_3$ will hold.

\noindent {\bf Case 10.2:} Suppose the resulting sequence is $x\ \
b_2\ \ a_1\ \ a_2\ \ b_3\ \ b_1$. The next triple is $(b_2, a_2,
b_3)$.

\noindent {\bf Case 10.3:} Suppose the resulting sequence is $x\ \
b_2\ \ a_1\ \ a_2\ \ b_1\ \ b_3$. The next triple is $(b_2, a_2,
b_3)$.

Note that in the last two subcases, if the next triple is used to
choose $a_3 > b_3$ that sees $a_2$ and misses $b_2$, then $a_1 <
a_3$ will hold.

\noindent
{\bf Case 11:} $x\ \ a_2\ \ b_2\ \ a_1\ \ b_1$\\
Use the triple $(x, a_2, b_1)$ to choose $a_3 > b_1$ that sees $a_2$
and misses $x$. Use Lemma~\ref{lem:extend} to get the extended bull
$B(x, a_3, b_2)$. The resulting sequence is $x\ \ a_2\ \ b_2\ \ a_1\
\ b_1\ \ a_3$ and the next triple is $(a_2, b_2, a_3)$. Note that if
this triple is used to choose $b_3 > a_3$ that sees $b_2$ and misses
$a_2$, then $b_1 < b_3$ will hold.

\noindent
{\bf Case 12:} $x\ \ b_2\ \ a_2\ \ a_1\ \ b_1$\\
Use the triple $(x, b_2, a_1)$ and choose $b_3 > a_1$ that sees
$b_2$ and misses $x$. Apply Lemma~\ref{lem:extend} to get the
extended bull $B(x, a_2, b_3)$. The resulting sequence is either $x\
\ b_2\ \ a_2\ \ a_1\ \ b_3\ \ b_1$ or $x\ \ b_2\ \ a_2\ \ a_1\ \
b_1\ \ b_3$. In each case, the next triple is $(b_2, a_2, b_3)$.
Again, if this triple is used to choose $a_3 > b_3$ that sees $a_2$
and misses $b_2$, then $a_1 < a_3$ will hold.

Note that any next triple is either of the form $(a_{i-1}, b_j,
a_i)$ or $(b_{j-1}, a_i, b_j)$; without loss of generality, assume
that it is the latter and we already have the extended bull $B(x,
a_i, b_j)$. Apply $(P_*)$ to the triple $(b_{j-1}, a_i, b_j)$ to
choose the rightmost vertex $a_{i+1}$ such that $a_{i+1} > b_j$ and
$a_{i+1}$ sees $a_i$ but misses $b_{j-1}$.

Observe that either $a_3$ is brought in as the rightmost vertex that
sees $a_2$ and misses $x$ or it is the rightmost vertex that sees
$a_2$ and misses some $b_r$. In the former case, $a_3$ satisfies the
conditions of the Lemma~\ref{lem:extend}. Otherwise, consider the
general case. Assume $a_p$, $3 \leq p \leq i$, is brought in as the
rightmost vertex that sees $a_{p-1}$ and misses some $b_q$. Given
that $B(x, a_i, b_j)$ is an extended bull, $a_p$ clearly misses
every $b_q$. Therefore, it can be deduced that $a_p$ is indeed the
rightmost vertex that sees $a_{p-1}$ and misses every $b_q$. Thus,
every $a_p$ satisfies conditions of Lemma~\ref{lem:extend}. By
symmetry, every $b_q$ also satisfies the conditions of
Lemma~\ref{lem:extend}.

Applying Lemma~\ref{lem:extend}, we get the larger extended bull
$B(x, a_{i+1}, b_j)$. As we now have $a_i < b_j < a_{i+1}$, we can
use $(a_i, b_j, a_{i+1})$ as the next triple to complete the
inductive argument. \hfill \qed

\begin{corollary} \label{cor:scheme}
Suppose $G$ is a (building, gem)-free graph. Then, every induced
subgraph $H$ of $G$ contains a vertex $v$ such that $v$ is not the
nose of any bull in $H$.
\end{corollary}

\begin{figure}[h]
\epsfxsize=8cm \centerline{\epsfbox{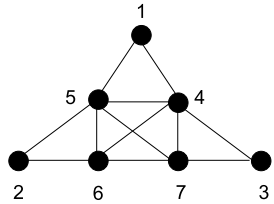}}
%
%
\vskip -5cm
\caption{A building-free graph with an LBFS ordering where the first
vertex is the nose of a bull.} \label{fig:risingsun}
\end{figure}

Figure~\ref{fig:risingsun} shows that without forbidding the gem,
Theorem~\ref{thm:scheme} cannot be proved. For a vertex $v$ in graph
$G$, let $D_2(v)$ denote the set of those vertices of $G$ that are
within distance two from $v$. The following theorem was proved in
\cite{DN}.

\begin{theorem} \label{thm:dh} \normalfont{\cite{DN}}
Graph $G$ is distance hereditary if and only if for every LBFS
ordering ${v_1}{v_2}\ldots{v_n}$ of $G$ and for each $v_i$,
$D_2(v_i)$ in $G[{v_i}, \ldots, {v_n}]$ does not contain a $P_4$.
\end{theorem}

If $v$ is a vertex of $G$ such that $D_2(v)$ does not contain a
$P_4$, then $v$ is not the nose of any bull in $G$. However, the
converse is not true; consider the graph constructed by starting
with the $P_4$ $abcd$ and adding vertex $v$ adjacent only to $a$ and
$c$. Therefore, the elimination scheme addressed in
Theorem~\ref{thm:scheme} is more general than the one given in
Theorem~\ref{thm:dh}.

\begin{figure}[h]
\epsfxsize=8cm \centerline{\epsfbox{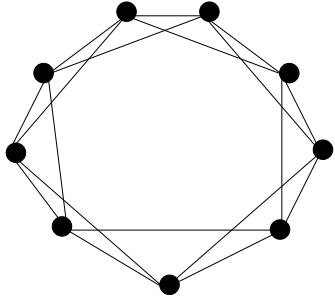}}
%
%
\vskip -4cm
\caption{In $(C_9)^2$ every vertex is nose of a bull.}
\label{fig:midbull}
\end{figure}

In view of Corollary~\ref{cor:scheme}, let $\mathcal{C}$ denote the
class of building-free graphs $G$ such that in every induced
subgraph $H$ of $G$ there exists a vertex that is not nose of any
bull in $H$. What can we say about the class $\mathcal{C}$? Every
strongly chordal graph belongs to the class $\mathcal{C}$. More
generally, every {\it strongly orderable graph} \cite{Dr} that does
not contain a building belongs to the class $\mathcal{C}$. Clearly,
any $k$-sun, $k \geq 4$, is a minimal forbidden subgraph for the
class $\mathcal{C}$. Conversely, what are minimal graphs in which
every vertex is nose of some bull? Figure~\ref{fig:midbull} is
$(C_9)^2$, which is minimal for the property. We note that this
graph contains buildings.

\section{Discussion}
We conclude the paper with some problems. A module $M$ in graph $G =
(V, E)$ is a set of vertices such that $M \not= \emptyset$, $M \not=
V$, and for every vertex $x \in V - M$, either $x$ sees every vertex
in $M$ or $x$ misses every vertex in $M$. The use of modules in the
design of efficient algorithms is a well-researched subject
\cite{B}. It was proved in \cite{HK} that every hhd-free graph
contains a module or a simplicial vertex; this result has been used
to design an efficient recognition algorithm for the class of
hhd-free graphs \cite{HS}. In view of this, is it true that every
hhds-free graph contains a module or a simple vertex?

Given that a sun in a graph can be found efficiently when the graph
belongs to some perfect graph classes, what is the complexity of
finding a sun in a weakly chordal graph? A graph is {\it weakly
chordal} if neither the graph nor its complement contains any holes.
More generally, what is the complexity of finding a sun in a perfect
graph?

\end{document}